\documentclass[prl,superscriptaddress,twocolumn]{revtex4}

\usepackage{amsfonts,amssymb,amsmath}
\usepackage[]{graphics,graphicx,epsfig}
\usepackage{amsthm}

\def\identity{\leavevmode\hbox{\small1\kern-3.8pt\normalsize1}}

\renewcommand{\epsilon}{\varepsilon}

\begin{document}
\title{Practically state-independent test of contextuality with 9 observables}

\date{\today}

\begin{abstract}

\end{abstract}

%\preprint{APS/123-QED}

\author{P. \surname{Kurzy\'nski}}
\email{cqtpkk@nus.edu.sg}
\affiliation{Centre for Quantum Technologies, National University of Singapore, 3 Science Drive 2, 117543 Singapore, Singapore}
\affiliation{Faculty of Physics, Adam Mickiewicz University, Umultowska 85, 61-614 Pozna\'{n}, Poland} 
\author{D. \surname{Kaszlikowski}}
\email{phykd@nus.edu.sg}
\affiliation{Centre for Quantum Technologies, National University of Singapore, 3 Science Drive 2, 117543 Singapore, Singapore}
\affiliation{Department of Physics, National University of Singapore, 2 Science Drive 3, 117542 Singapore, Singapore}

%\date{\today}% It is always \today, today,
             %  but any date may be explicitly specified

\begin{abstract}
We propose a test of quantum contextuality for a single three-level system that uses nine projective measurements. It has a form of an inequality that has to be satisfied by any non-contextual theory and which is violated by any quantum state, except for the maximally mixed one. Due to the fact that there is only a single state that does not exhibit contextuality, it is natural to ask what the difference between state-independent tests and the one proposed here is. 
\end{abstract}

%\pacs{}% PACS, the Physics and Astronomy
                             % Classification Scheme.
%\keywords{Suggested keywords}%Use showkeys class option if keyword
                              %display desired
\maketitle

The phenomenon of quantum contextuality is manifested by the fact that the outcomes of measurements may depend on the choice of what is co-measured at the same time. The smallest quantum system exhibiting contextuality is a single qutrit. In this case one needs at least five measurements to construct a test capable of revealing contextuality for some class of quantum states (state-dependent test) \cite{Kl}, or thirteen measurements for a test that reveals contextuality for any quantum state (state-independent test) \cite{YuOh}. In this work we propose a test that uses nine measurements and reveals contextuality for all, but a maximally mixed state.

The original proof of quantum contextuality due to Kochen and Specker \cite{KS} involved 117 projective measurements in the real three-dimensional space. Since then the number of necessary measurements drastically decreased and only recently Yu and Oh \cite{YuOh} proposed a test to reveal contextuality for all quantum states that involves 13 projectors. Later, it was proven by Cabello \cite{C1} that one cannot obtain a state-independent proof of contextuality using less than 13 measurements. 

The 13-projector test is not the minimal one. It is enough to use only four measurements \cite{CHSH}, or in case of a qutrit five measurements \cite{Kl}, to reveal contextuality. However, if one uses less than 13 measurements then the test becomes state-dependent. The most acclaimed state dependent tests are due to Clauser-Horne-Shimony-Holt (CHSH) \cite{CHSH} and to Klyachko-Can-Binicioglu-Shumovsky (KCBS) \cite{Kl}. The former tests nonlocality of a bipartite system, which is a special form of contextuality, and reveals it for a subset of entangled states. The latter tests a single three-level quantum system which in this case exhibits contextuality for a special class of states.

Any test constructed with less than 13 measurements would not reveal contextuality for some subset of quantum states. It may seem that the more projectors we use, the smaller the size of this subset is. Here, we show that already for 9 projective measurements this subset consists of only one element, namely the maximally mixed state. As a consequence, it is natural to ask what the meaning of state-independent tests of quantum contextuality is. We observe that the relation between state-independent and state-dependent tests might be due to the interplay between prior knowledge of the state and the memory needed to store the measurement data.

The paper is organized as follows. First, we define nine projective measurements and discuss their orthogonality relations. We use them to construct an inequality which has to be obeyed by any non-contextual theory. Next, we show that the maximally mixed state saturates this inequality and that any other state violates it. Finally, we discuss the properties of our inequality and its relation to other known tests of quantum contextuality. 

\begin{figure}
\scalebox{0.5}
{\includegraphics{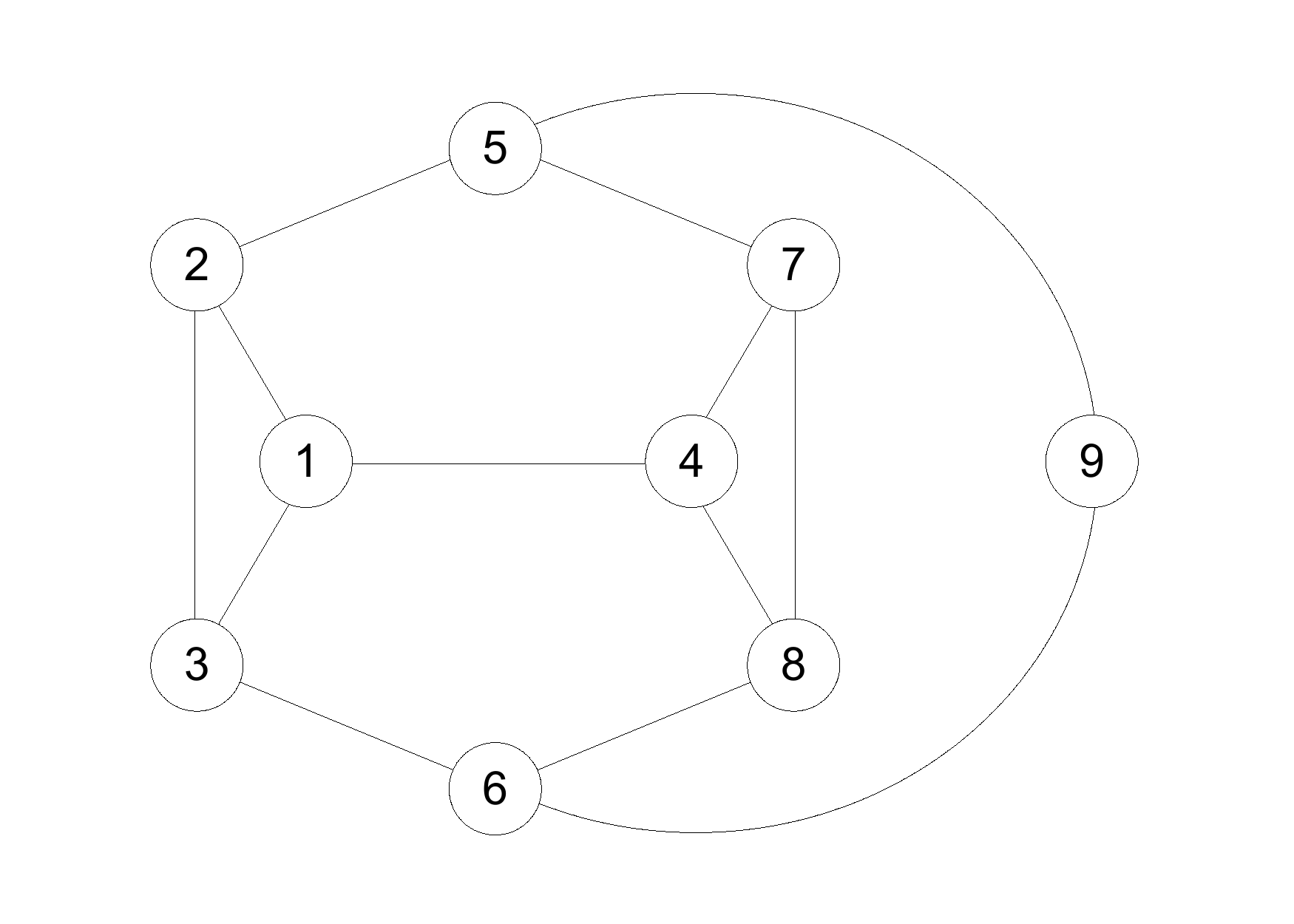}}
\vspace{-0.5 cm}
\caption{\label{f1} Plot of the graph $G$ representing orthogonality relations between nine projectors.}
\end{figure}

Let us consider nine three-dimensional projectors
\begin{equation}
\Pi_i=|i\rangle\langle i|. \nonumber
\end{equation}
They project onto the following real vectors
\begin{eqnarray}
|1\rangle&=&(1,0,0)^T, \nonumber \\
|2\rangle&=&(0,1,0)^T, \nonumber \\
|3\rangle&=&(0,0,1)^T, \nonumber \\
|4\rangle&=&1/\sqrt{2}(0,1,-1)^T, \nonumber \\
|5\rangle&=&1/\sqrt{3}(1,0,-\sqrt{2})^T, \nonumber \\
|6\rangle&=&1/\sqrt{3}(1,\sqrt{2},0)^T, \nonumber \\
|7\rangle&=&1/2(\sqrt{2},1,1)^T, \nonumber \\
|8\rangle&=&1/2(\sqrt{2},-1,-1)^T, \nonumber \\
|9\rangle&=&1/2(\sqrt{2},-1,1)^T. \nonumber
\end{eqnarray}
The orthogonality relation between these vectors is presented in the Fig. \ref{f1}, where the vectors are represented by vertices and the orthogonality relation is represented by edges.  
The graph $G$, except for vertex number 9, corresponds to the original building block of Kochen and Specker \cite{KS} and consists of two joint pentagons that are primitives for contextuality in dimension three \cite{Kl,Us}.

In any realistic theory one assigns definite values to projectors, either 1 or 0. A non-contextual theory corresponds to a probability distribution over all possible realistic configurations, therefore it is important to study what configurations are possible. It was argued in Ref. \cite{CSW} that non-contextual bound $\cal{C}$ for the inequality based on orthogonality graphs
\begin{equation}
\sum_{i\in V(G)} \langle \Pi_i \rangle \leq {\cal{C}} \nonumber
\end{equation}
is given by the graph independence number, i.e., the maximal number of vertices that are not connected. The intuitive reason for this is that one wants to assign 1 to as many vertices as possible, however due to orthogonality relation it is not possible for the two neighboring vertices to be assigned 1 at the same time. The set of vertices that are assigned 1 has to form an independent set. The independence number of the graph in the Fig. \ref{f1} is three, therefore the inequality yields 
\begin{equation}\label{iq1}
\sum_{i=1}^9 \langle \Pi_i \rangle \leq 3.
\end{equation}

One can rewrite the above inequality in terms of correlations between $\pm 1$ operators
\begin{equation}
A_i=\openone - 2\Pi_i. \nonumber
\end{equation}
After simple algebra and with the help of the following relation 
\begin{equation}
A_i A_j=\openone - 2\Pi_i - 2\Pi_j \nonumber
\end{equation}
one obtains an equivalent form of (\ref{iq1})
\begin{equation}\label{iq2}
\sum_{(i,j)\in E(G)} \langle A_i A_j \rangle + \langle A_9\rangle \geq -4.
\end{equation}
The reason that the single term $\langle A_9\rangle$ appears is that the 9'th vertex in $G$ is the only vertex whose degree (the number of connections) is two, not three.

Before we discuss the violation of inequalities (\ref{iq1}) and (\ref{iq2}), let us observe that the maximally mixed state $\rho=\openone/3$ saturates them. In case of (\ref{iq1}) we have $\langle \Pi_i \rangle=1/3$ for all the projectors and since we sum over nine projectors, the sum is three. In case of (\ref{iq2}) $\langle A_9 \rangle=1/3$ and $\langle A_i A_j \rangle = -1/3$ for all pairs $(i,j)$. Therefore, the left hand side equals to -4 since the total number of edges in $G$ is 13.

Now, let us show that for any quantum state that is different from the maximally mixed one, there exists an inequality (\ref{iq1}), or equivalently (\ref{iq2}), that is violated. First, we note that the left hand side of (\ref{iq1}) can be expressed as
\begin{equation}
\sum_{i=1}^9 \langle \Pi_i \rangle=\langle C \rangle, \nonumber
\end{equation}
where $C$ is a $3\times 3$ matrix which is a sum of nine projectors. The eigenvalues of $C$ are 
\begin{equation}
\lambda_1=\frac{10}{3},~~\lambda_2=3,~~\lambda_3=\frac{8}{3}. \nonumber
\end{equation}
Any $3\times 3$ density matrix $\rho$ can be expressed in some diagonal basis as
\begin{equation}
\rho=p|\psi_1\rangle\langle\psi_1|+q|\psi_2\rangle\langle\psi_2|+r|\psi_3\rangle\langle\psi_3|, \nonumber
\end{equation}
where 
\begin{equation}
p \geq q \geq r. \nonumber
\end{equation}
One can always find an operator $C$ for which
\begin{equation}\label{e3}
\langle C \rangle =\text{Tr}(\rho C)=\frac{1}{3}(10p + 9q +8r)=3+\frac{p-r}{3},
\end{equation} 
since $q=1-p-r$. The only situation when (\ref{e3}) equals 3 is when $p=r=1/3$, which happens only when $\rho$ is the maximally mixed state. In all other cases one obtains violation.

At this point it is natural to ask why the maximally mixed state is so different from the other states and what the difference between inequalities (\ref{iq1}) and (\ref{iq2}) and the state-independent inequalities like the ones discussed in \cite{YuOh,C2,C3} is. For most state-dependent tests the state preparation procedure is of great importance. For example, in the case of CHSH inequality one has to be certain that the state is entangled and is pure enough in order to allow for the violation. However, in our case the state preparation procedure seems to be of very little importance. Namely, it is enough to have at most a minor control over the source just to ensure that the state is not maximally mixed. In particular, the important class of physically accessible finite temperature thermal states is contextual with respect to our test.

To understand better the relation between our test and the truly state-independent tests, let us consider the following scenario. Imagine that there are two parties which we call Alice and Bob (by convention). Alice has access to a source of particles that posses a three-level degree of freedom (say spin 1), which she can control to some extend. She prepares an ensemble of particles that is represented by a density matrix $\rho$. Next, the whole ensemble is passed to Bob. His goal is to choose a proper inequality to test and to violate it on the ensemble he received from Alice. We are going to consider three cases.

First, let us assume that Bob is capable of performing a state-independent test like the one proposed in \cite{YuOh}. Moreover, we also assume that he does not know anything about the Alice's preparation procedure. From his point of view the state of the ensemble is completely random, i.e., maximally mixed $\rho=\openone/3$. However, since the test reveals contextuality for any state, he does not need to know anything about the ensemble to violate the corresponding inequality. In this case it seems that the violation is due to the properties of the measurements only.

Next, consider the similar scenario, but this time Bob is allowed to perform less than 13 measurements. This might be due to limited memory resources that are needed to store the outcomes of measurements. If he can store at least outcomes of nine measurements, then he can perform our test. However, since he does not have any information about the source and the Alice's preparation, he can at most saturate the inequality, but not violate it.

Finally, we assume that Bob has some prior information about the ensemble. From his point of view its state is given by $\rho_B \neq \openone/3$. This information can be measured by simple function of von Neumann entropy $S(\rho)=-\text{Tr}(\rho\log \rho)$
\begin{equation}
\eta = S(\openone/3) - S(\rho_B) = \log 3 - S(\rho_B). \nonumber
\end{equation}
In this case, whenever $\eta \neq 0$ Bob can always find nine measurements that will reveal contextuality of $\rho_B$, which is a simple consequence of Eq. (\ref{e3}). Interestingly, if Bob were given more information about the ensemble he would be able to use even less measurements. For example, enough information would be sufficient to use only five measurements and to reveal contextuality via KCBS test \cite{Kl} or the entropic one considered in \cite{Us,FC}. 

To conclude, we speculate that the relation between state-dependent and state-independent tests of quantum contextuality might be interpreted as an interplay between prior knowledge of the state of the ensemble and the memory needed to store the outcomes of measurements. It would be interesting to further explore this relation.

This work was supported by the National Research Foundation and Ministry of Education in Singapore. PK was also supported by Foundation for Polish Science. We would like to thank Ravishankar Ramanathan for stimulating discussions.

\end{document}